\documentclass[conference,letterpaper]{IEEEtran}
\usepackage{pifont}
\usepackage{braket}
\usepackage{amsfonts, amsmath}
\usepackage{cite}
\ifCLASSINFOpdf
  \usepackage[pdftex]{graphicx}
\else
\fi
\usepackage{amsmath}
\usepackage{mhchem}

\ifCLASSOPTIONcompsoc
  \usepackage[caption=false,font=normalsize,labelfont=sf,textfont=sf]{subfig}
\else
  \usepackage[caption=false,font=footnotesize]{subfig}
\fi
\newcommand{\TextSub}[1]{\ensuremath{_{\mbox{\tiny{#1}}}}}  

\begin{document}
%
\title{Electric-field Inputs for Molecular Quantum-dot Cellular Automata Circuits}

\author{\IEEEauthorblockN{Enrique Blair}
\IEEEauthorblockA{Electrical and Computer Engineering Department\\
Baylor University\\
Waco, Texas 76798, United States of America\\
Email: Enrique\_Blair@baylor.edu}}


%


\maketitle

\begin{abstract}
Quantum-dot cellular automata (QCA) is a low-power, non-von-Neumann, general-purpose paradigm for classical computing using transistor-free logic. An elementary QCA device called a ``cell'' is made from a system of coupled quantum dots with a few mobile charges. The cell's charge configuration encodes a bit, and quantum charge tunneling within a cell enables device switching. Arrays of cells networked locally via the electrostatic field form QCA circuits, which mix logic, memory and interconnect. A molecular QCA implementation promises ultra-high device densities, high switching speeds, and room-temperature operation. We propose a novel approach to the technical challenge of transducing bits from larger conventional devices to nanoscale QCA molecules. This signal transduction begins with lithographically-formed electrodes placed on the device plane. A voltage applied across these electrodes establishes an in-plane electric field, which selects a bit packet on a large QCA input circuit. A typical QCA binary wire may be used to transmit a smaller bit packet of a size more suitable for processing from this input to other QCA circuitry. In contrast to previous concepts for bit inputs to molecular QCA, this approach requires neither special QCA cells with fixed states nor nanoelectrodes which establish fields with single-electron specificity. A brief overview of the QCA paradigm is given. Proof-of-principle simulation results are shown, demonstrating the input concept in circuits made from two-dot QCA cells. Importantly, this concept for bit inputs to molecular QCA may enable solutions to or provide insights into other challenges to the realization of molecular QCA, such as the demonstration of molecular device switching, the read-out of molecular QCA states, and the layout of molecular QCA circuits.
\end{abstract}


%
\IEEEpeerreviewmaketitle

\section{Introduction}

The need for an replacement for the prevalent transistor-based computing paradigm is becoming increasingly urgent. The decades-long trend of scaling transistors has been fruitful, steadily increasing device densities. However, the approach of fundamental physical limits on photolithography imposes limits on device scaling and makes further progress costly and increasingly difficult. Quantum effects become more significant and can disrupt transistor operation, and chief among the problems of nano-scale transistors is vast power dissipation \cite{FrankCMOSscalingLimints}. This power dissipation is responsible not only for the 2003 plateau in commercial computer clock speeds, but also the vast amounts of energy consumed and dissipated by computers. Information and communication technologies now consume a significant and growing percentage of the global electrical power: one estimate was that power consumed by information and computing technologies would grow from 10\% of global electric power production in 2010 to  30-50\% by 2030 \cite{ChallengesPowerConsumption}; however, this estimate was made before the significant increase in cryptocurrency mining activities in 2017.

A departure from transistor-based computing, quantum-dot cellular automata (QCA) is a non-von-Neumann paradigm for general-purpose, classical computing, which was designed to leverage quantum phenomena and allow energy-efficient devices \cite{LentTougawArchitecture,LentTougawPorodBernstein:1993}. QCA may be implemented using molecules, which promise ultra-high device densities and THz-speed-or-better switching speeds at room temperature \cite{LentScience2000,molecularQCAelectronTransfer}.

Here, a brief overview of QCA is given, with a focus on the molecular implementation. A solution is presented for the technical challenge of writing input bits to molecular QCA circuits. This solution consists of lithographically-formed input electrodes and molecular QCA circuits responsive to an applied electrostatic field. An earlier approach to molecular QCA inputs \cite{WalusQCAInput} requires the use of complementary sets of fixed QCA cells, which likely would be molecules of a  species different from that which comprises the rest of the QCA circuitry. On the other hand, the approach presented here does not require special fixed molecules. The approach here also is related to work by Pulimeno, \textit{et al} \cite{2012molecularWire}, but relaxes an assumption found therein about input electrodes with single-molecule specificity. Quantum mechanical simulations are used to demonstrate a viable method for bit write-in to molecular QCA circuits.

This theoretical work plays an important part in bringing molecular QCA closer to realization. Theoretical and simulated demonstrations of solutions to paradigmatic challenges pave the way for creative and compelling experimental work and collaborative partnerships between theorists and experimentalists. Theory and simulation not only can demonstrate the viability and promise of molecular QCA, but also accelerate the realization of molecular QCA computation devices by predictive modeling. No molecules have been synthesized and tested as QCA devices yet, and the design-synthesis-testing process can take several person-years of effort. Relevant quantum phenomena such as electron transport \cite{molecularQCAelectronTransfer} and environmental effects\textemdash including decoherence \cite{BlairLentJAP2013QuantumDecoherence,2017RamseyBlair}, the loss of entanglement \cite{2018LossOfEntanglement}, and power dissipation \cite{2011QCAEnergyRecovery}\textemdash  may be modeled for candidate QCA molecules, and their performance as QCA devices may be predicted \cite{2018MoleculeDesign}. Thus, theory and models can play an important role in closing a theoretical design loop for molecules and save significant resources by enabling researchers to avoid synthesizing and testing unoptimized molecules.

\section{Overview of QCA}
\begin{figure}[htbp] 
   \centering
   \includegraphics[height=1in]{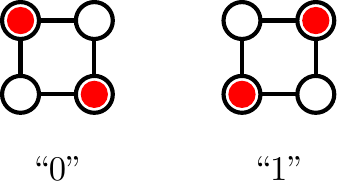} 
   \caption{The charge configuration of a four-dot QCA cell provides a bit. Two electrons (red discs) reside in a system of four quantum dots (black circles). Lines connecting dots indicate quantum tunneling paths for the electrons. Two localized localized electronic states provide a ``0'' state and a ``1'' state.}
   \label{fig:FourDotCellStates}
\end{figure}

In QCA, the elementary device is a \textit{cell}, a system of coupled of quantum dots which provide charge localization sites for a few mobile charges. The configuration of charge on these dots encodes a bit, and device switching occurs via the quantum tunneling of charge between the dots. An example of this is shown in Fig.\ \ref{fig:FourDotCellStates}. Here, two mobile electrons on a system of four quantum dots arranged in a square provides two information-bearing, localized electronic configurations. The quantum tunneling of charge between dots enables device switching.

\begin{figure}[htbp] 
   \centering
   \includegraphics[width=0.485\textwidth]{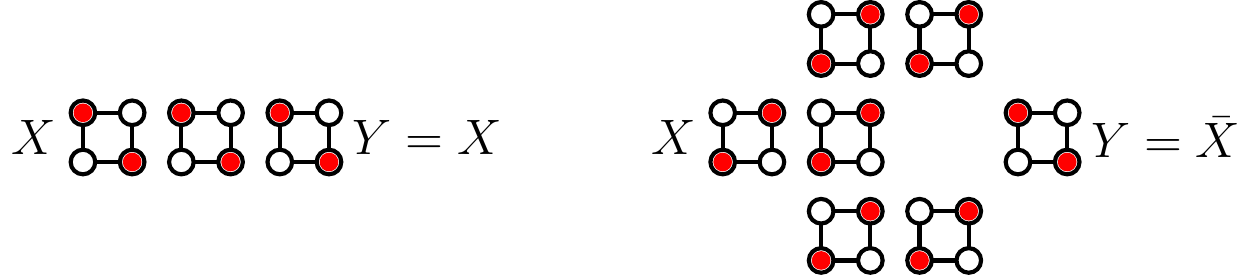} 
   \caption{Arrays of four-dot QCA cells form simple devices. (a) The states of cells arranged in a row align through Coulomb coupling, forming a binary wire. The wire's output $Y$ matches its input $X$. (b) Diagonal (next-nearest-neighbor) coupling inverts the bit so that the wire output $Y=\bar{X}$ is the logical complement of $X$. Since the diagonal coupling is weaker than the direct, nearest-neighbor coupling, two such diagonal couplings to the output cell enhance the coupling.}
   \label{fig:WireAndInverter}
\end{figure}

Circuits are constructed by forming arrays of QCA cells networked by the local electrostatic field. Coulomb repulsion causes neighboring cells to align, and this is the basis for the binary wire [see Fig.\ \ref{fig:WireAndInverter}(a)]. Bit inversion can also be achieved using a next-nearest-neighbor interaction [see Fig.\ \ref{fig:WireAndInverter}(b)]. Additionally, QCA cells can form a majority gate \cite{Snider1999,Imre2006} (not shown). The inverter and the majority gate provide a logically-complete set of QCA gates, from which more complex, general-purpose systems have been designed, including a Simple-12 processor \cite{Simple12Conf}.

QCA has several implementations. The earliest QCA devices were built from metallic dots on an insulator \cite{Orlov1997,Snider1998SemiconSciTech,Orlov199BinaryWire,QCASixDotLatch2001}. QCA devices also were implemented using semiconductor quantum dots \cite{Gardelis2003,Smith2003}, and more recently at the atomic scale using dangling bonds on a hydrogen-passivated Si surface \cite{WolkowQCA_Silicon}. While the metallic and semiconductor implementations require cryogenic operating temperatures, the atomic-scale implementation provides bit energies robust at room temperature by virtue of its small scale and strong Coulomb interactions. In the molecular QCA implementation of particular interest here, redox centers on mixed-valence molecules provide quantum dots \cite{QCA_at_molecular_scale,mQCAeFieldSwitch2003,molecular_QCA}. This implementation also provides small device dimensions and supports room-temperature operation.

\begin{figure}[htbp] 
   \centering
   \includegraphics[width=0.485\textwidth]{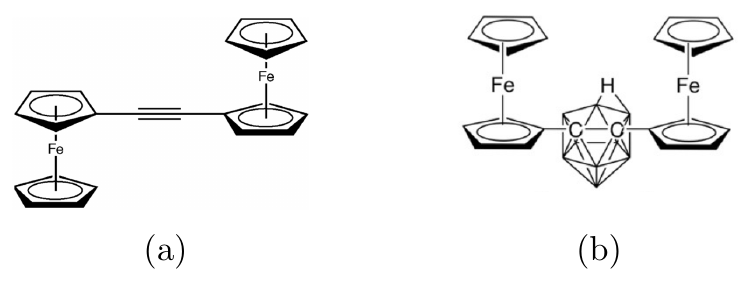} 
   \caption{Some molecules have been studied as QCA candidates. (a) The two Fe centers of diferrocenyl acetylene each provide one quantum dot. (b) The self-doping zwitterionic nido-carborane molecule was designed as a QCA candidate. Two Fe centers and one carborane cage of provide a system of three quantum dots for one mobile hole and one fixed electron (charges not shown).}
   \label{fig:Molecules}
\end{figure}

Work on the realization of QCA molecules is ongoing. Charge localization has been observed in some molecules \cite{mQCA_ChargeLocalization,mQCA_STMImaging3Dot}, and efforts in physics \cite{2018MoleculeDesign}, chemistry \cite{molecularQCAelectronTransfer}, and physical chemistry \cite{mQCA_PhysChem_SelfTrapping} are focused on the design, synthesis, and testing of molecules. Diferrocenyl acetylene [DFA, see Fig.\ \ref{fig:Molecules}(a)] is a mixed-valence species considered as a QCA candidate \cite{DFA}; and, zwitterionic nido-carborane [see Fig.\ \ref{fig:Molecules}(b)] has been designed and synthesized as a self-doping QCA candidate molecule \cite{Christie15}. Zwitterionic nido-carborane is significant because it eliminates the need for a counterion to neutralize each molecule.

While the concept of information processing has been established in QCA, important challenges exist in path to the realization of molecular QCA computation. Major challenges include the lack of experimentally-demonstrated technologies for (1) the layout of molecular QCA circuits; (2) the transduction of bits from macroscopic devices to molecular QCA; and (3) the readout of molecular bits to macroscopic systems.

Some solutions have been proposed, but further development is required before experimental demonstrations are achieved. DNA tiles have been proposed for self-assembled molecular QCA circuit layout \cite{QCADNATiles}. A previous proposal for molecular QCA write-in exists, but it requires the use of fixed-state QCA molecules as well as mastery over the layout of individual molecules \cite{WalusQCAInput}. Another previous proposal for inputs to molecular QCA circuits assumed an electric field with single-molecule specificity \cite{2012molecularWire}. For detecting the state of QCA molecules, single-electron transistors (SETs) may play a key role: SETs have demonstrated sensitivity to sub-nanometer displacements of individual electrons\cite{SETSensitivitySnider}.

In this paper, we explore bit write-in to QCA circuits using neither fixed-state molecules nor fields having single-molecule specificity. Here, a system is devised and modeled for using an applied input electric field to select the state of several molecules in a QCA circuit. Such a field can be realized using electrodes feasible using electron-beam lithography. The requirement for single-molecule specificity in the field (and thus the electrodes) is relaxed, and fields are considered in two idealized limits: (1) a uniform input field is applied to all molecules in the circuit, and (2) a non-uniform field comprised of two uniform regions with an abrupt transition between the two regions. The former case corresponds to the limit in which electrodes are much larger than the molecules, and the latter case corresponds to a case with lithographically-feasible electrodes but negligible fringing fields.

This work is related to\textemdash but distinct from\textemdash work by Pulimeno, \textit{et al}, in which \textit{ab initio} calculations were performed for a clocked QCA shift register of bisferrocene molecules terminated by a single input molecule controlled by nanoelectrodes \cite{2012molecularWire}. Here, we treat the molecules at a more abstract level and focus on the electric field itself. Thus, this work is not specific to any particular molecule. Also, the assumption of single-molecule specificity in the field and the electrodes is relaxed here, as already mentioned. Finally, in this case, unclocked, two-dot molecules are considered, as they provide minimal complexity and lighter calculations than do three-dot molecules, but still enable the study of electric-field inputs.

Methods for molecular QCA inputs like the one proposed here are important because they could lead to experimental demonstrations of controlled switching in molecular QCA, which are detectable by an STM tip, Raman spectroscopy, or an SET-based readout system. Also, this work may lead to insights into and place helpful constraints on technologies for molecular circuit layout.

\section{Model}

We develop a model of molecular QCA circuits immersed in an electric field used to select the state of the circuit. These circuits will be termed ``field-input circuits.'' The circuits studied are comprised of two-dot QCA molecules, as they provide the minimum complexity required to explore the problem of electric-field inputs to QCA circuits, with mathematical simplicity and a minimal state space. Electrodes patterned on the device surface may be used to create the input electric fields. A uniform electric field $\vec{E}$ is used at first, and then the circuit response is improved by applying a inhomogeneous field $\vec{E} (\vec{r})$. The uniform $\vec{E}$ corresponds to the limit of electrodes much larger than the circuits, and the inhomogeneous field $\vec{E} (\vec{r})$ represents a case in which a field of some specificity can be engineered.

\subsection{Individual QCA Cell in the Presence of a Driver and a Biasing Field $\vec{E}$}

\begin{figure}[htbp] 
   \centering
   \includegraphics[height=1.25in]{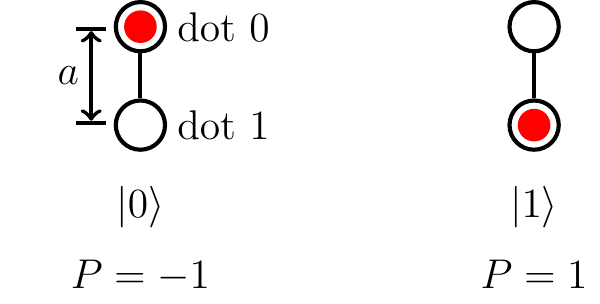} 
   \caption{One mobile electron on two coupled quantum dots provides two charge-localized states of a two-dot QCA cell. The dots are separated by a distance $a$, and the line connecting line between dots indicates the tunneling path. The charge configuration is encoded on a single number $P \in [-1, 1]$, the cell's polarization.}
   \label{fig:TwoDotStates}
\end{figure}

Fig.\ \ref{fig:TwoDotStates} shows the states of a two-dot cell, the simplest QCA device, which may function as half of a four-dot cell. The two localized states of one mobile electron of charge $-q_e$ on a coupled pair of quantum dots provides binary states $\ket{0}$ and $\ket{1}$ ($q_e$ is the fundamental charge).  These states form an ordered computational basis $\mathcal{B} = \{ \ket{1}, \ket{0}\}$ used in this discussion. The electronic configuration of the two-dot cell may be encoded in a single number $P$, the cell polarization, defined as the expectation value of the $\hat{\sigma}_z$ operator:
\begin{equation}
P \equiv \braket{\hat{\sigma}_z}\; ,
\end{equation}
where $\hat{\sigma}_z$ is one of the Pauli operators $\{\hat{\sigma}_x, \hat{\sigma}_y, \hat{\sigma}_z\}$:
\begin{eqnarray}
\hat{\sigma}_x & = & \ket{0}\bra{1} + \ket{1}\bra{0} \; , \nonumber \\
\hat{\sigma}_y & = & -i \left( \ket{0}\bra{1} - \ket{1}\bra{0} \right) \; , \; \mbox{and} \nonumber \\
\hat{\sigma}_z & = &  \ket{1}\bra{1} - \ket{0}\bra{0}  \; .
\end{eqnarray}
Not pictured in Fig.\ \ref{fig:TwoDotStates} is the fixed neutralizing charge, which endows a QCA cell with a net-zero charge, preventing tremendous repulsive electrostatic forces between cells in a circuit. Here, we assume that one neutralizing charge of $+q_e$ is split evenly between the two dots and resides at the center of each dot. All charges are treated as point charges.   

\begin{figure}[htbp] 
   \centering
   \includegraphics[width=0.485\textwidth]{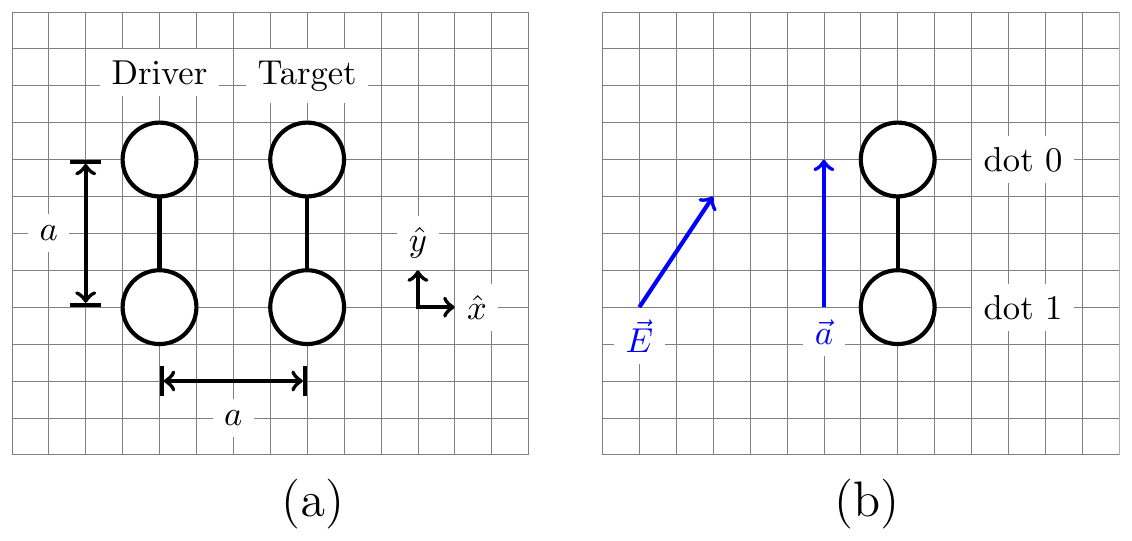} 
   \caption{Neighboring (``driver'') cells and applied (``input'') electric fields can influence a target molecule. (a) The charge configuration of a driver establishes an electrostatic electric field that biases the state of the target cell. (b) Interaction between the electrostatic field and a target double-dot QCA molecule may be modeled as field-dipole interaction. Here, the vector $\vec{a}$ is a displacement vector from dot 1 to dot 0. The electrostatic electric field $\vec{E}$ provides a input-field detuning $\Delta_E = -q_e \vec{E} \cdot \vec{a}$ between the states $\ket{0}$ and $\ket{1}$.}
   \label{fig:DrivenCellWithInputField}
\end{figure}

In the presence of a driver molecule and zero applied field, as shown in Fig.\ \ref{fig:DrivenCellWithInputField}(a), the target molecule has a Hamiltonian given by
\begin{equation}
\hat{H} = -\gamma \hat{\sigma}_x +  \frac{\Delta}{2} \hat{\sigma}_z \; .
\end{equation}
Here, $\gamma$ is a tunneling energy between the two basis states of $\mathcal{B}$; and $\Delta$ is the cell detuning, the difference between occupation energies of the computational basis states.
\begin{equation}
\Delta = \braket{1| \hat{H}| 1} - \braket{0| \hat{H} | 0} \; .
\end{equation}
To include the interaction of a molecule with the electrostatic field $\vec{E}$, as in Fig.\ \ref{fig:DrivenCellWithInputField}(b), we include the field biasing $\Delta_E$ between the two states $\ket{0}$ and $\ket{1}$, given by
\begin{equation}
\Delta_E = -q_e \vec{E} \cdot \vec{a} \; .
\end{equation}
Here, $\vec{a}$ is a displacement vector from dot 1 to dot 0. The negative sign takes into account the sign of the mobile electron. Thus, the complete Hamiltonian for a target cell\textemdash including driver interaction and field interaction\textemdash is
\begin{equation}
\hat{H} = -\gamma \hat{\sigma}_x +  \frac{\Delta + \Delta_E}{2} \hat{\sigma}_z \; .
\label{eqn:HamiltonianFieldAndDriver}
\end{equation}

\begin{figure}[htbp] 
   \centering
   \includegraphics[width=0.485\textwidth]{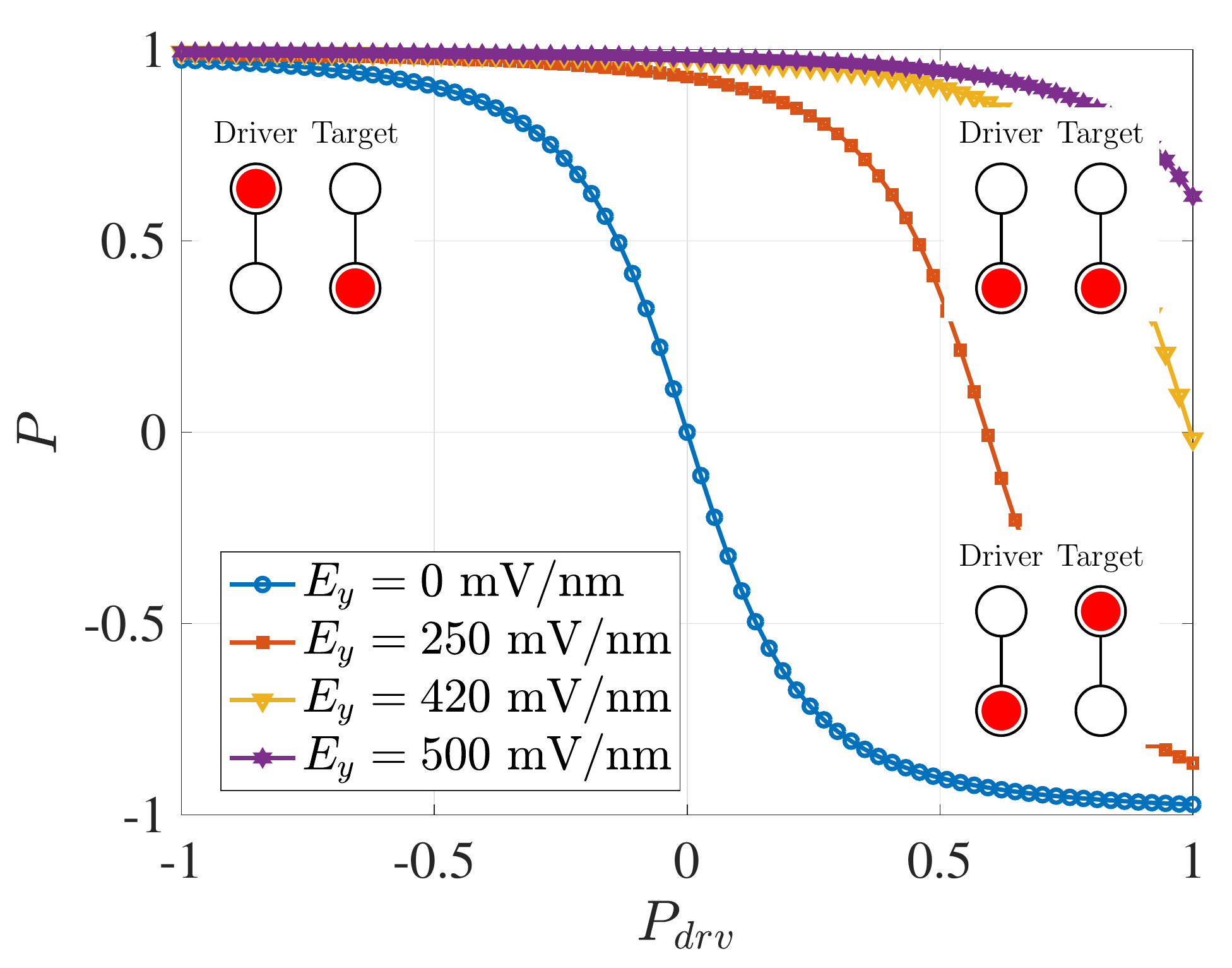} 
   \caption{A target cell responds to a polarized driver cell in the presence of an electric field. The target cell's response to the driver is shown for different values of the field strength $E_y$. In the absence of the field ($E_y = 0$), the ground state of the target molecule is anti-aligned with that of the driver. A sufficiently-strong field ($E_y>420~\mbox{meV}$) can prevent the driver from switching the target.}
   \label{fig:DriverVersusField}
\end{figure}

Fig.\ \ref{fig:DriverVersusField} shows the response of the target cell to both the single driver molecule and the input field $\vec{E}$. Since the cell's $\vec{a}$ has only a $y$-component, only the $y$-component of the field interacts with the cell. When $E_y= 0$, the cell polarization response $P$ to the driver's polarization $P_{drv}$ has odd symmetry about the origin. That is, the target cell tends to anti-align with the driver in the absence of an applied input field, and the driver effectively selects the state of the target cell.  As $E_y$ is increased from zero, the state $\ket{1}$ ($P\rightarrow 1$) is preferred. The presence of an applied field can affect the state of the target molecule, even selecting the target molecule's state in the absence of a driver.

The kink energy $E_k$ provides a meaningful energy scale for this system. A kink energy is the cost of a full bit flip in the target cell when the driver cell is fully polarized ($P_{drv} = \pm 1$) in the absence of any biasing field ($\vec{E}\cdot \vec{a} = 0$).  Equivalently, $E_k$ is the detuning of the target cell when the driver cell is fully polarized ($P_{drv} = 1$) and  $\vec{E}\cdot \vec{a} = 0$:
\begin{equation}
E_k \equiv \braket{1 | \hat{H} |1} - \braket{0 | \hat{H} |0} \quad \mbox{with} \quad P_{drv} = 1, \vec{E} \cdot \vec{a} = 0 \; .
\end{equation}
The geometry alone of the QCA cell determines $E_k$, and it may be considered as a function of only the cell characteristic length $a$. Here, we use $a=1~\mbox{nm}$, so $E_k = 420.7~\mbox{meV}$.

When the field is strong enough to overcome the kink energy ($|\Delta_E| = q_e \vec{E} \cdot \vec{a} \geq E_k$), a fully-polarized driver is insufficient to flip the sign of $P$. Thus,
\begin{equation}
E_o = \left| \frac{E_k}{q_e a} \right|
\end{equation}
is the strength of the component of $\vec{E}$ parallel to $\vec{a}$ needed to overcome biasing from a single, fully-polarized driver cell. Fig.\ \ref{fig:DriverVersusField} provides an example of this in the $E_y = 420~\mbox{mV/nm}$. Here,  $\Delta _E \simeq E_k$ approaches a field strength which prevents the driver from selecting $P < 0$ on the target cell. The $E_y = 500~\mbox{mV/nm}$ is strong enough to ``lock'' the target cell in a $P>0$ state.

\subsection{Molecular QCA Circuits Immersed in an Electric Field} \label{subsect:CircuitModel}

Next, we develop a model for an $N$-cell circuit of two-dot molecules immersed in a uniform field. The $2^N$-dimensional Hilbert space $\mathbb{H}$ for the circuit is the direct product of 2-dimenstional Hilbert spaces $\mathbb{H}_k$ ($k=1, 2, \ldots , N$)  for each of the $N$ cells: $\mathbb{H} = \mathbb{H}_N \otimes \cdots \otimes \mathbb{H}_1$. Thus, the basis states $\mathcal{B}_k = \{ \ket{0_k},  \ket{1_k} \}$ for all cells may be used to form a set of $2^N$ basis states $\{ \ket{\vec{m}_p} \}$ for the circuit:
\begin{eqnarray}
\ket{\vec{m}_p} & = &  \ket{m_N m_{N-1} \cdots m_2 m_1 }  \nonumber \\
 & = & \ket{m_N } \ket{ m_{N-1}} \cdots \ket{m_2} \ket{ m_1 } \; ,
\end{eqnarray}
with $m_k \in \, \{ 0, 1 \}$. Here, $p = \sum_{k=1}^N m_k 2^{k-1} \in \{0, 1, 2, \ldots, 2^N-1\}$ is an integer uniquely labeling each basis state.

Since the objective here is to develop QCA input circuits that do not need fixed cells or driver cells, the Hamiltonian for the $k$-th cell in isolation is written without the detuning $\Delta$, which is an applied potential from driver molecules:
\begin{equation}
\hat{H}_k = -\gamma \hat{\sigma}_x + \frac{\Delta_E}{2} \hat{\sigma}_z\; .
\end{equation}
Here, the field-interaction term $\Delta_E \hat{\sigma}_z/2$ is considered for each cell.

The Hamiltonian $\hat{H}$ for an $N$-cell circuit, then, includes $\hat{H}_k$ for each cell, as well as an interaction term $\hat{H} \TextSub{int}$:
\begin{equation}
\hat{H} = \sum_{k=1}^N \hat{H}_k + \hat{H} \TextSub{int} \; .
\end{equation}
The interaction term $\hat{H} \TextSub{int}$ is diagonal in the basis $\{ \ket{\vec{m}_p} \}$, and each diagonal element $\braket{\vec{m}_p | \hat{H} \TextSub{int} |\vec{m}_p}$ is found by summing the electrostatic energy of interaction between all pairs of cells in the states $\{ \ket{m_k} \}$ as determined by $\vec{m}_p$. To find the state of a circuit immersed in electric field $\vec{E}(\vec{r})$, we find its ground state.

\section{Results}

The model developed in Section \ref{subsect:CircuitModel} is used to simulate QCA circuits immersed in a uniform electric field $\vec{E}$. No driver cells are present. We consider longitudinal arrays of QCA cells, which respond effectively to the input field; transverse arrays of QCA cells, which form binary wires; and a transverse array coupled to a longitudinal array. Finally, results are provided for an ideal inhomogeneous field $\vec{E} (\vec{r})$ designed to improve the performance of the molecular QCA input circuits.

\subsection{Longitudinal Arrays}

\begin{figure}[htbp] 
   \centering
   \includegraphics[width=0.485\textwidth]{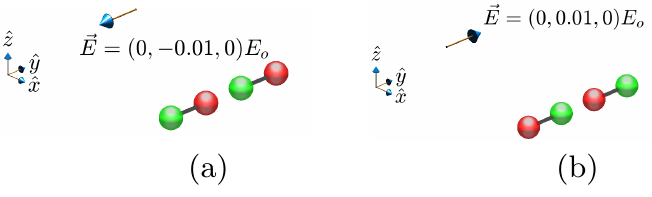} 
   \caption{Cells aligned along their longitudinal axes are highly sensitive to the parallel component of the uniform, applied electrostatic field $\vec{E}$. Here, paired colored spheres represent molecular quantum dots, and the connecting bar indicates a intramolecular tunneling path. A red color indicates the presence of the mobile electron, and the green color indicates the uncovered positive, fixed neutralizing charge. A relatively weak $y$-component may be used to switch the state of this longitudinal array from $\ket{0}\ket{0}$ in subfigure (a) to $\ket{1}\ket{1}$ in subfigure (b). Here, $a=1~\mbox{nm}$ and $\gamma=1~\mbox{meV}$.}
   \label{fig:FieldCTRLLongitudinalN2}
\end{figure}

Cells aligned along an axis passing through the dots of each cell are termed longitudinal arrays, as the axis of alignment coincides with the longitudinal axis of each cell. A two-cell longitudinal array is seen in Fig.\ \ref{fig:FieldCTRLLongitudinalN2}.
A longitudinal array of QCA cells is sensitive to the field component in its axis of alignment. In Fig.\  \ref{fig:FieldCTRLLongitudinalN2}, $E_y$, the $y$-component of $\vec{E}$, controls the state of the cells in the array. The cells in the array interact with the field via a dipole interaction and are highly sensitive to the sign of $E_y$. A very small field is sufficient to select a bit on the cells of the longitudinal array. Bit selection and switching is demonstrated in Fig.\  \ref{fig:FieldCTRLLongitudinalN2} with $|E_y| = E_o/100$. This sensitivity enables longitudinal arrays to form the field-sensitive element of a molecular QCA field-input circuit. In this calculation and other uniform-field calculations, the choice of $a=1~\mbox{nm}$ is inspired by the length scale of actual molecular QCA candidates. A low tunneling energy $\gamma=1~\mbox{meV}$ is used to simplify the analysis of the circuits.

\subsection{Transverse Arrays} \label{subsect:Results_TxArr}

\begin{figure}[htbp] 
   \centering
   \includegraphics[width=0.485\textwidth]{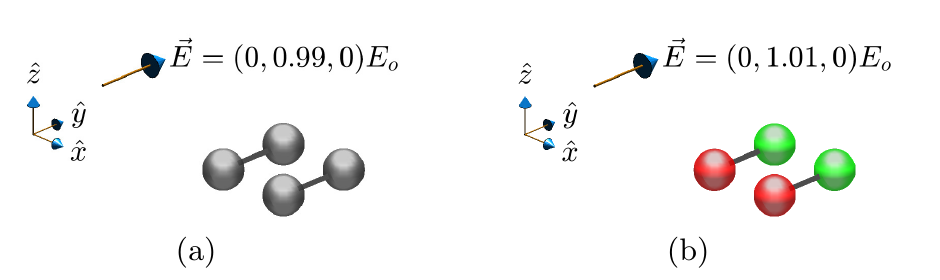} 
   \caption{Cells aligned along in a direction ($\pm \hat{x}$) transverse to their longitudinal axes ($\pm \hat{y}$) demonstrate poor QCA circuit behavior in the presence of the $y$-component of the field. Subfigure (a): when the field is insufficient to introduce a kink in the two-cell configuration, the two states $\ket{01}$ and $\ket{10}$ are degenerate, and the circuit  ground state is a linear combination of these localized states. Thus, the two cells are depolarized (gray spheres). Subfigure (b): when the field is sufficient to induce a kink, the dipole interaction with the field overcomes the intercellular dipole interaction, and all cells align with the field. Neither of these behaviors are conducive circuit function as a binary wire. For these results, $a=1~\mbox{nm}$ and $\gamma=1~\mbox{meV}$ were chosen.}
   \label{fig:FieldCTRLBroadsideN2}
\end{figure}

In a transverse array, the axis of alignment between constituent cells runs transverse to the longitudinal axis of the individual members. A simple transverse array with $N=2$ cells is shown in Fig.\ \ref{fig:FieldCTRLBroadsideN2}. Here the axis of alignment between cells is in the $\pm \hat{x}$ direction, but the cells each have their vectors $\vec{a}$ aligned in the $\pm \hat{y}$ direction. Thus, only $E_y$, the $y$-component of $\vec{E}$ can affect the state of these cells. Transverse arrays of cells typically are considered binary wires, in which each two-dot molecule is a half of a four-dot cell. A functioning wire will consist of two-dot cells alternating in state, or four-dot cells copying their state to the next four-dot cell in the line.

\begin{figure}[htbp] 
   \centering
   \includegraphics[width=0.485\textwidth]{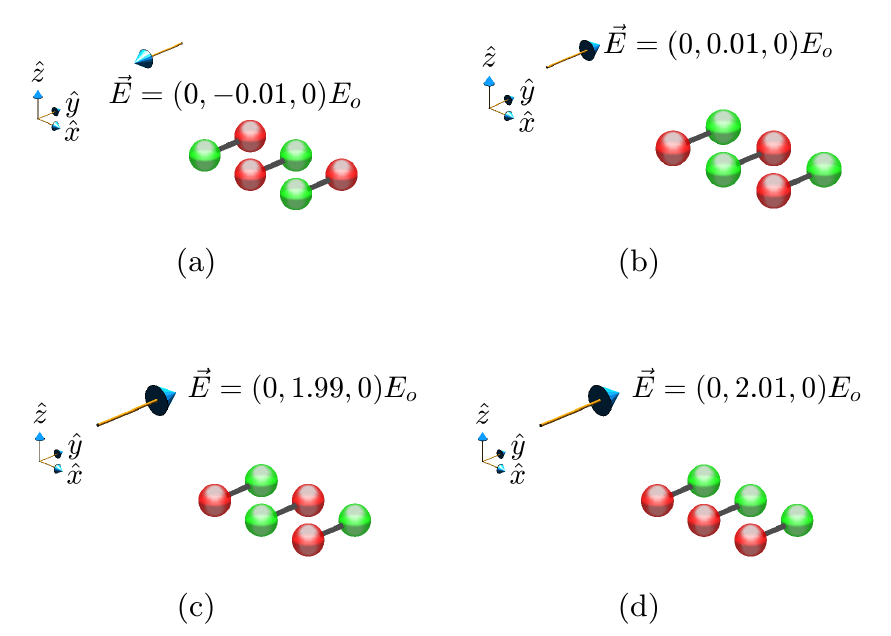} 
   \caption{A transverse array of three QCA cells ($N=3$) demonstrates odd-parity effects. Figures (a) and (b) show that unlike the $N=2$ transverse array, a weak $E_y$ polarizes the system. Figure (c) and (d) show that $E_y$ must be greater than $2 E_o$ so that it can force two kinks into the system (one on either side of the middle cell).  Here, $a=1~\mbox{nm}$ and $\gamma=10~\mbox{meV}$.}
   \label{fig:FieldCTRLBroadsideN3}
\end{figure}

Transverse arrays exhibit parity-dependent behaviors that affect their suitability as an input-field sensitive element. First, consider the $N=2$ case of Fig.\ \ref{fig:FieldCTRLBroadsideN2}. This is the simplest case of even parity ($N$ is even) in the transverse array. In the limit of $\gamma=0$, when $|E_y| < E_o$, the field does not polarize the cells; but, if $|E_y| > E_o$, and both cells polarize in alignment with the field, and a binary wire state is not achieved. On the other hand, if another cell is added for odd parity ($N=3$), a relatively weak $E_y$ is sufficient to polarize the cells and select a binary wire configuration (see Fig.\ \ref{fig:FieldCTRLBroadsideN3}). A strong $E_y$ will cause failure in the odd-parity binary wire, also [see Fig.\ \ref{fig:FieldCTRLBroadsideN3}(d)].

\begin{figure}[htbp] 
   \centering
   \includegraphics[width=0.485\textwidth]{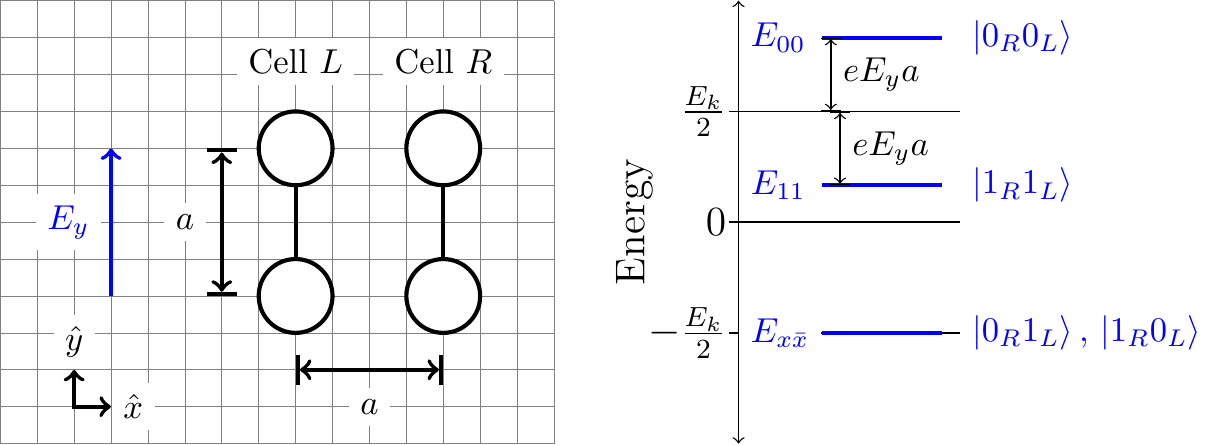} 
   \caption{Energy levels for the two-cell system immersed in an electrostatic field.}
   \label{fig:EnergyLevelsN2Broadside}
\end{figure}

The result for the $N=2$ case may be understood simply by counting field-charge interactions and electrostatic interactions in the limit of $\gamma \rightarrow 0$ with the help of Fig.\ \ref{fig:EnergyLevelsN2Broadside}, in which the two cells are labeled ``left'' ($L$) and ``right''  ($R$). Here, we will take the zero of intercellular interaction energy to be the average of the kinked and unkinked energies of interaction. Thus, a single kink (states $\ket{0_R0_L}$ or $\ket{1_R1_L}$) adds $E_k/2$ of intercellular interaction energy to the total circuit energy, and an anti-aligned state ($\ket{1_R0_L}$ or $\ket{0_R1_L}$) adds $-E_k/2$. The zero of field interaction will be taken as the energy of the electron midway between the zero dot and the one dot. For an applied field $E_y$, a cell in the state $\ket{1}$ contributes $-q_e E_y a/2$, and a cell in the state $\ket{0}$ contributes $q_e E_y a/2$ to total system energy.  Thus, the states $\ket{0_R1_L}$ and $\ket{1_R0_L}$ have energy $E_{01} = E_{10} = -E_k /2$. The state $\ket{0_R0_L}$ has one kink and two electrons, each in the $\ket{0}$ state, for a state energy $E_{00} = E_k/2 + q_e E_y a$. Similarly, the state $\ket{1_R1_L}$ has occupation energy $E_{11} = E_k/2 - q_e E_y a$. These energy levels are plotted in Fig.\ \ref{fig:EnergyLevelsN2Broadside}. For $E_y < E_o$, the states $\ket{01}$ and $\ket{10}$\textemdash now with subscripts $R$ and $L$ omitted\textemdash are degenerate minimum-energy states. The ground state will be a linear superposition of the low-energy states: $(1/\sqrt{2}) \left(\ket{01} + \ket{10}\right)$. Here, both cells are depolarized, as seen in the simulated result of Fig.\ \ref{fig:FieldCTRLBroadsideN2}(a). When $E_y > E_o$, on the other hand, $q_e E_y a > E_k$, and $E_{11}$ drops below $E_{x\bar{x}} = E_{01} = E_{10}$, so that $\ket{11}$ is the ground state, and both cells align with the field, as in the result of Fig.\ \ref{fig:FieldCTRLBroadsideN2}(b). Thus, the kink energy $E_k$ causes the system to interact weakly as a quadrupole with field $E_y < E_o$, until a strong field $E_y > E_o$ injects a kink of energy and interacts with each cell as a dipole.

When a third double-dot molecule is added, as in Fig.\ \ref{fig:FieldCTRLBroadsideN3}, the states $\ket{010}$ and $\ket{101}$ (subscripts are omitted) are accessible with a field $|E_y| < 2 E_o$ [see Figs.\ \ref{fig:FieldCTRLBroadsideN3}(a)-(c)]. Here, the interaction with the field may be thought of as a field-dipole interaction and a field-quadrupole interaction. The field readily interacts with the dipole, driving it to the state $\ket{1}$. The dipole then then biases one of the localized quadrupole electronic configurations $\ket{01}$ or $\ket{10}$ of the remaining two-molecule pair. Thus, the desirable alternating states characteristic of binary wire arise on the system of two-dot cells. On the other hand, a field $|E_y| > 2 E_o$ is strong enough to cause two kinks between the outer cells and the middle cell. Thus, all the cells polarize fully to the $\ket{1}$ state, for a global state $\ket{111}$ [see Fig.\ \ref{fig:FieldCTRLBroadsideN3}(d)].

\subsection{QCA Field-input Arrays}

\begin{figure}[htbp] 
   \centering
   \includegraphics[width=0.485\textwidth]{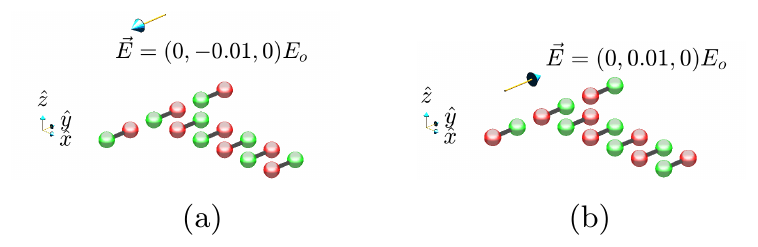} 
   \caption{A longitudinal array is coupled to a transverse array to form an input-field-controlled QCA circuit. Here, the longitudinal array is responsive to the field and biases a state in the transverse array. In contrast to the result of the even-parity transverse array, even a weak field ($E_y \ll E_o$) can polarize the transverse array.  Here, $a=1~\mbox{nm}$ and $\gamma=10~\mbox{meV}$.}
   \label{fig:FieldCTRLCircuitN8}
\end{figure}

A longitudinal array may be coupled to a transverse array to form the proposed QCA field-input array, as shown in Fig.\ \ref{fig:FieldCTRLCircuitN8}. This system uses the longitudinal array as the $E_y$-sensitive component. The state of the longitudinal array then biases a polarized state of the transverse array, selecting an appropriate binary-wire response in the transverse array. This then, can couple a bit from the input to molecular QCA logic circuits for processing. As shown in Fig.\ \ref{fig:FieldCTRLCircuitN8}, a weak field ($E_y < E_o$) is sufficient to select the state in the longitudinal array and achieve a polarized response from the transverse array. Switching is demonstrated by reversing the sign of $E_y$.

\begin{figure}[htbp] 
   \centering
   \includegraphics[width=0.485\textwidth]{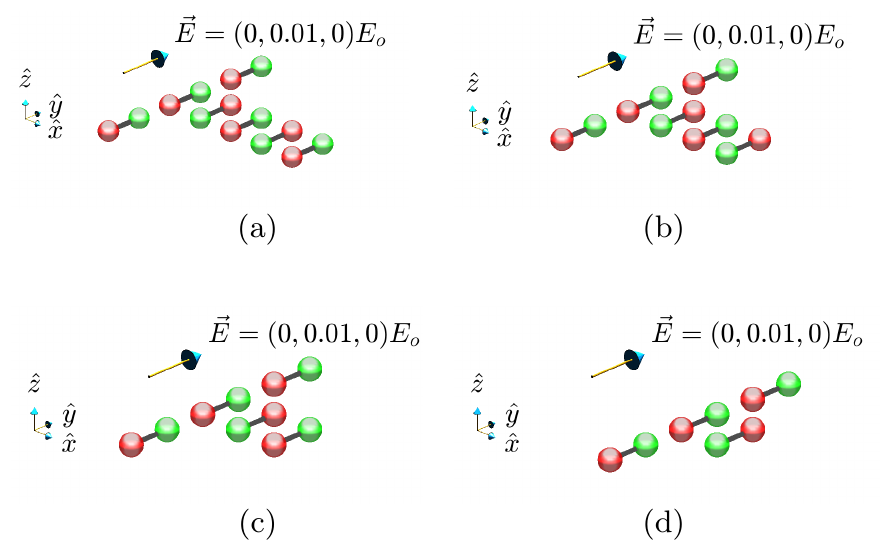} 
   \caption{Molecular QCA field-input circuits demonstrate a robustness against the parity-dependent effects observed in Sect.\ \ref{subsect:Results_TxArr}. Here, the transverse array length is varied from $N=4$ [subfigure (a)] down to $N=1$ [subfigure (e)], and in each case, the result remains consistent with the bit selected by the input field $\vec{E}$. Here, $a=1~\mbox{nm}$, and $\gamma=10~\mbox{meV}$.}
   \label{fig:FieldCTRLCircuitNvaried}
\end{figure}

The QCA field-input array also exhibits robustness against the parity-dependent effects seen in Section \ref{subsect:Results_TxArr}. In Fig.\ \ref{fig:FieldCTRLCircuitNvaried}, the applied field $\vec{E}$ remains constant, and the number of cells in the transverse array is varied. In each case, the longitudinal array responds to the field as desired, and the longitudinal array selects the appropriate binary wire state for the transverse array.

\begin{figure}[htbp] 
   \centering
   \includegraphics[width=0.485\textwidth]{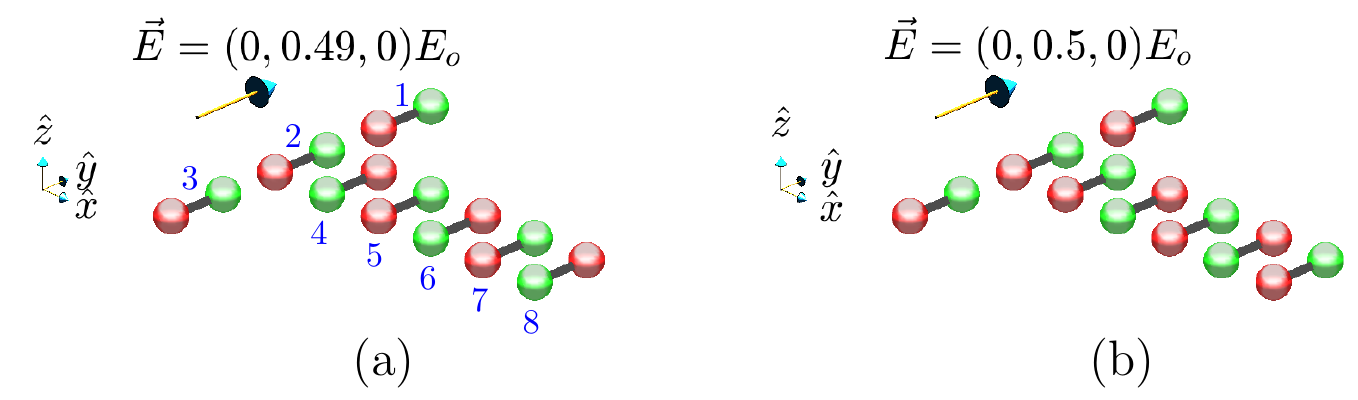} 
   \caption{A molecular QCA field-input circuit fails under a strong input field. Subfigure (a): when $|E_y| < E_o/2$, the transverse array couples properly to the longitudinal array. The output $P_8$, the polarization of cell 8 matches the input polarization selected on the cells in the longitudinal array (cells 1-3). Subfigure (b): when $|E_y| \geq \sim E_o/2$, the interaction of cell 4 with cells 1 and 3 enables the field to cause a kink between cells 2 and 4, and the output bit (on cell 8) flips, now matching the bit selected on the cells of the longitudinal array. Here, $a=1~\mbox{nm}$ and $\gamma=10~\mbox{meV}$.}
   \label{fig:FieldCTRLCircuitFailure}
\end{figure}

A QCA field-input array can fail under high uniform fields. A vulnerability is shown in Fig.\ \ref{fig:FieldCTRLCircuitFailure}, where a field-input circuit is shown with eight cells. Cells 1-3 form the longitudinal array, and cells 4-8 comprise a transverse segment. When the uniform field $\vec{E}$ has $|E_y| < E_o/2$, the circuit operates as desired [see Fig.\ \ref{fig:FieldCTRLCircuitFailure}(a)]; but, when $|E_y| \geq E_o/2$, interactions of cell 4 with cells 1 and 3 allow the field to introduce a kink between cells 2 and 4, even when $|E_y| < E_o$. The failed state is seen in Fig.\ \ref{fig:FieldCTRLCircuitFailure}(b), and the output bit on cell 8 is the reverse of the resultant output bit for $|E_y| < E_o$.

\begin{figure}[htbp] 
   \centering
   \includegraphics[width=0.425\textwidth]{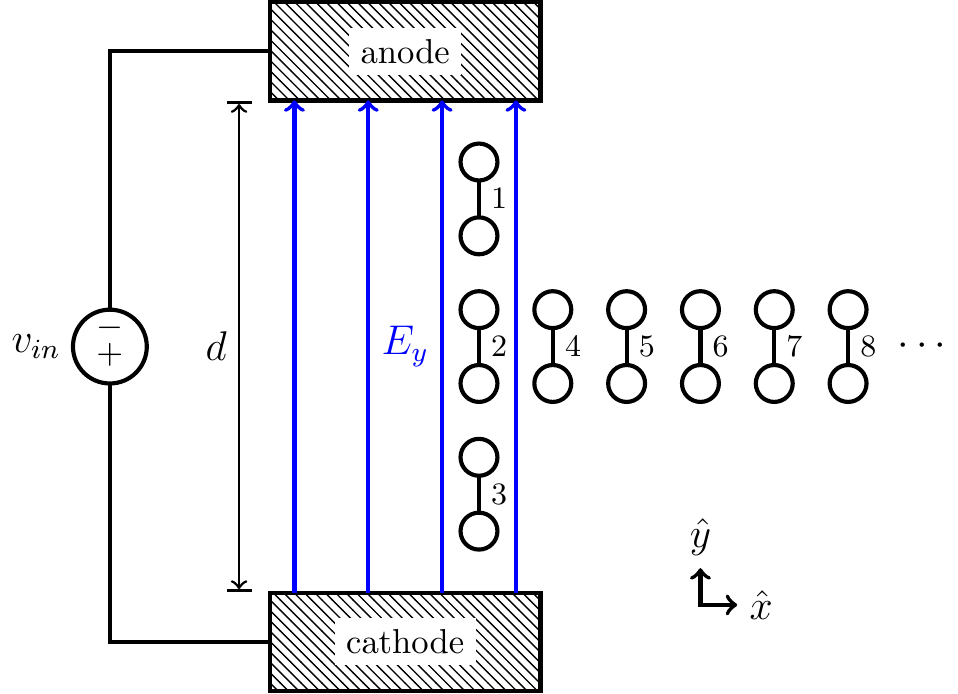} 
   \caption{An inhomogeneous electric field $\vec{E} (\vec{r})$ may be engineered to have minimal impact on the transverse section of the molecular QCA input circuit. Here, a voltage $v_{in}$ is applied to a pair of electrodes separated by distance $d$. The electrodes are designed to establish a uniform field ($E_y \hat{y}$, shown in blue lines) within the volume between them and minimal field outside that volume, where the transverse array (cells 4-8) lies. Thus, the input field directly affects only the longitudinal array (cells 1-3).}
   \label{fig:FieldCTRLIdeal}
\end{figure}

While the preceding calculations are for a uniform field $\vec{E}$, an inhomogeneous field $E(\vec{r})$ better-suited for a QCA input circuit may be designed. Such a field could be realized by patterning electrodes on the device plane in a manner that establishes the in-plane input field across the longitudinal array, but minimizes the in-plane field in the region of the transverse array. This is idealized and illustrated in Fig.\ \ref{fig:FieldCTRLIdeal}. Here, two electrodes with separation $d$ straddle the longitudinal array of cells 1-3. Voltage $v_{in}$ applied across the electrodes establishes the applied field, having a nonzero component only in the $\pm \hat{y}$ direction with strength $E_y = v_{in}/d$. Field fringing is ignored so that the $x$- and $y$-components of the field are zero outside of the volume between the electrodes. This eliminates the parity-dependent effects altogether from the transverse array (cells 4-8) as well as the kink seen in Fig.\ \ref{fig:FieldCTRLCircuitFailure}(b) at the coupling between the longitudinal and transverse segments for $E_y > 0.5 E_o$.

\begin{figure}[htbp] 
   \centering
   \includegraphics[width=0.475\textwidth,trim={0.25cm 0 0.25cm 0},clip]{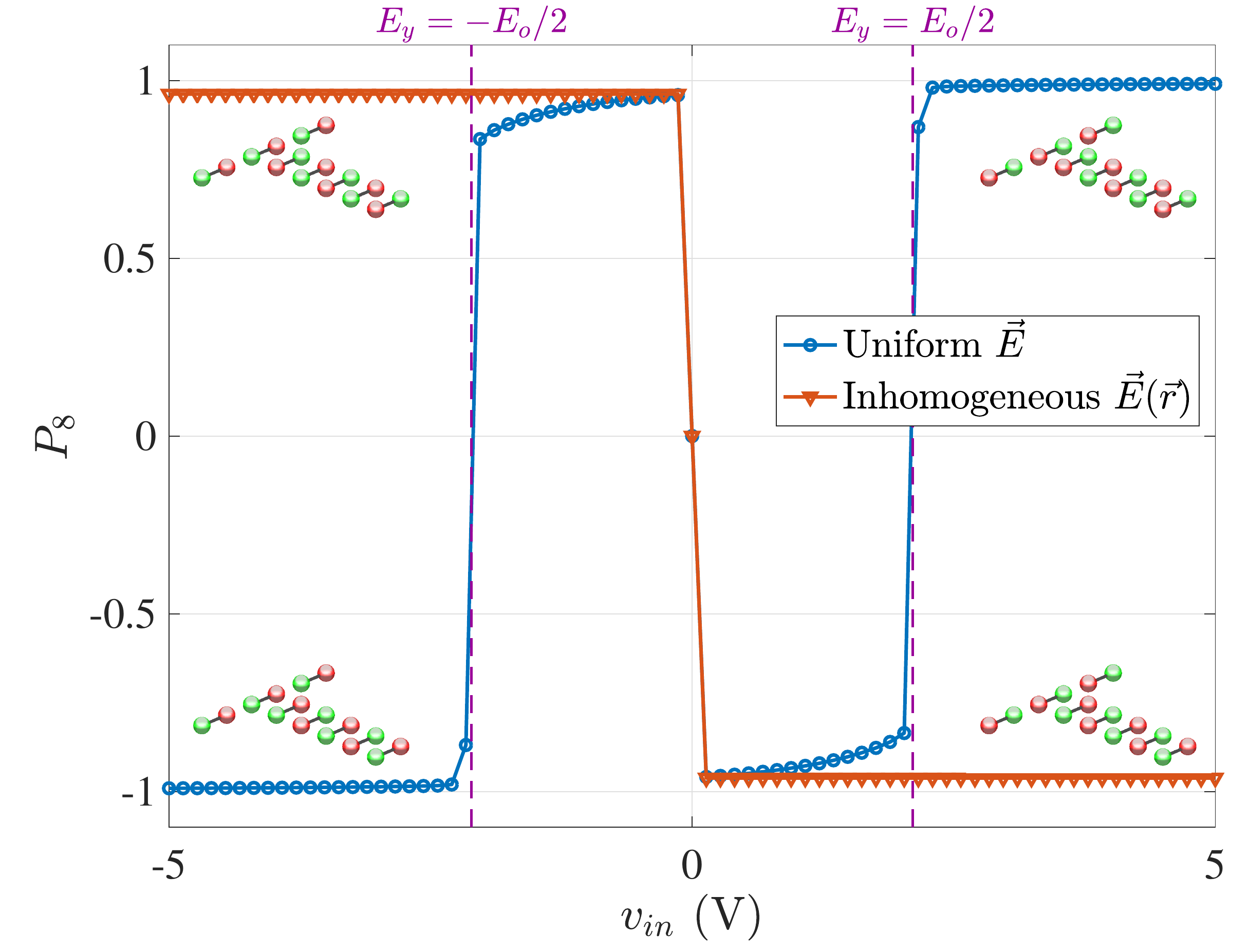} 
   \caption{An inhomogeneous electric field $\vec{E} (\vec{r})$ engineered to interact only with the longitudinal section provides a more effective input field than does the uniform $\vec{E}$. The absence of $\vec{E} (\vec{r})$ interaction with the transverse section prevents a kink from forming at the coupling between the longitudinal and transverse sections for $|E_y| \geq E_o/2$. For $\vec{E}(\vec{r})$, the output polarization $P_8$ of cell 8 always complements the signs of $v_{in}$ and $E_y$, regardless of the strength of the $E_y$. Under uniform $\vec{E}$, a kink forms, and the sign of $P_8$ flips when $|E_y|$ grows too large. The point of failure for the input circuit under the uniform field is marked by dashed lines for $|E_y| = E_o/2$. Here, $a=1~\mbox{nm}$, $\gamma=50~\mbox{meV}$, and $d=10~\mbox{nm}$.}
   \label{fig:FieldCTRLIdealCalc}
\end{figure}

Simulation results for the circuit and idealized $\vec{E} (\vec{r})$ of Fig.\ \ref{fig:FieldCTRLIdeal} are shown in Fig.\ \ref{fig:FieldCTRLIdealCalc} for various values of $v_{in}$. Parameters $a=1~\mbox{nm}$ and $\gamma=50~\mbox{meV}$ are chosen here because they are within the neighborhood of parameters characterizing the DFA molecule. Additionally, $d=10~\mbox{nm}$ was chosen because features of this scale are feasible through electron-beam lithography, and such a separation would accommodate a longitudinal array of length $N=3$ or $N=5$. A larger $d$ may be used, in which case a proportionally larger $v_{in}$ will be required to obtain the same response. Here, the polarization response $P_8$ of cell 8 (the output of the transverse section) is plotted in Fig.\ \ref{fig:FieldCTRLIdealCalc} versus $v_{in}$. $P_8$ is shown for both a uniform field $\vec{E}$ and the inhomogeneous $\vec{E} (\vec{r})$ constrained to the volume between the electrodes. The inhomogeneous $\vec{E} (\vec{r})$ yields a better result because it does not interact directly with the transverse segment, and the failure of Fig.\ \ref{fig:FieldCTRLCircuitFailure}(b) under the uniform $\vec{E}$ for $|E_y| \geq E_o/2$ is eliminated.  Indeed, it is preferable that the input field not interact with the binary wire.

\section{Conclusion}

We have shown that a longitudinal array of two-dot QCA cells coupled to a transverse array of QCA cells provides a field-input molecular QCA circuit. Here, a voltage applied across electrodes patterned on the device plane can create an input electric field that biases the state of a longitudinal molecular QCA array, which is coupled to a binary wire. This enables the input bit to be transmitted to QCA logic circuits. The input system proposed here does not require a separate species of fixed-polarization molecules in addition to the molecules comprising the computational logic; nor does it require electrodes that create fields with single-molecule specificity. If electrodes are much larger than the molecules such that the electrodes immerse the circuit in a single, uniform field, it remains possible to select an input without disrupting molecular wire functionality if the input field remains sufficiently weak. A more desirable case with improved circuit operation applies the input field only to input portions of the circuitry and suppresses the field in regions where binary wire behaviors are desired. 

This work demonstrates the proof of principle for this type of QCA input system in a minimal model, but more detailed work is merited. Non-ideal effects such as fringing fields and their impact on circuit operation also should be explored. The $z$-component of the field $\vec{E} (\vec{r}, t)$ may be used to control clocked molecular QCA systems \cite{BlairLentArchitecture}, and this may be leveraged to further minimize the impact of the input field on the binary wires (transverse section). Additionally, it is desirable to have experimental validation of these QCA input systems.

The molecular QCA input systems proposed here are of strategic importance in realizing molecular QCA computation. This theoretical work may lead directly to an experimental demonstration of controlled switching in individual QCA molecules, which can enable experimental demonstration of molecular QCA bit readout systems. This work also may inform desirable features of molecular QCA circuit layout solutions. By addressing the important technical challenge of using microscopic devices for inputting bits on nano-scale molecular QCA circuits, this work serves as a stepping stone in the path to the realization of low-power, high-speed general-purpose computing based on molecular QCA.

\section*{Acknowledgment}

This research was supported under a new-faculty-start-up grant from Baylor University. The author thanks C.S.\ Lent and G.L.\ Snider of the University of Notre Dame for insightful and engaging discussion.



%
\bibliographystyle{IEEEabrv}
\bibliography{QCAFieldControl.bib}

%
%

\end{document}